\documentclass{siuro250211}
\usepackage{graphicx,epstopdf}
\usepackage{float}
\usepackage[caption=false]{subfig}
\usepackage{mathtools}
\usepackage{amsfonts}
\usepackage{subcaption}
\usepackage{booktabs}
\usepackage{xcolor}
\usepackage{placeins}
\usepackage{cleveref}
\graphicspath{{figures/}}
\usepackage{titlesec}
\titlespacing*{\subsection}{0pt}{1ex plus 0.2ex minus 0.2ex}{0.5ex plus 0.1ex}

\title{Predicting COVID-19 Prevalence Using Wastewater RNA Surveillance: A Semi-Supervised Learning Approach with Temporal Feature Trust\thanks{Submitted to the editors DATE.}}

\author{
Yifei Chen\thanks{Mount Holyoke College, South Hadley, MA (\email{chen239y@mtholyoke.edu})}
\and
Eric Liang\thanks{Groton School, Groton, MA (\email{eliang26@groton.org})}
}

\dedication{\small\textit{Project advisor: Yao Li\thanks{University of Massachusetts Amherst, 710 N Pleasant St, Amherst, MA, 01002 (\email{liyao@umass.edu}).}}}

\headers{Predicting COVID-19 Using Wastewater RNA}{Yifei Chen and Eric Liang}

\begin{document}

\maketitle

\begin{abstract}
As COVID-19 transitions into an endemic disease that remains constantly present in the population at a stable level, monitoring its prevalence without invasive measures becomes increasingly important. In this paper, we present a deep neural network estimator for the COVID-19 daily case count based on wastewater surveillance data and other confounding factors. This work builds upon \cite{jiang2024artificial}, which connects the COVID-19 case counts with  testing data collected early in the pandemic. Using the COVID-19 testing data and the wastewater surveillance data during the period when both data were highly reliable, one can train an artificial neural network that learns the nonlinear relation between the COVID-19 daily case count and the wastewater viral RNA concentration. From a machine learning perspective, the main challenge lies in addressing temporal feature reliability, as the training data has different reliability over different time periods. 
\end{abstract}



\section{Introduction}\label{sec:introduction}
After the initial pandemic that caused millions of deaths and trillions in economic losses, COVID-19 has remained a constant threat to human society, particularly for high-risk groups. Therefore, it is crucial to accurately monitor COVID-19 prevalence without taking invasive measures like lockdowns and mass testing. All infected individuals, whether symptomatic or asymptomatic, shed the virus in their bodily waste, which eventually flows into wastewater systems. Therefore, the wastewater viral RNA becomes a novel surveillance tool to track community-level infections \cite{daughton2020wastewater, hillary2020wastewater, mcmahan2021covid}. 

The main goal of this paper is to find the hidden connection between the wastewater viral RNA surveillance data and the numerical count of the COVID-19 daily new infections. This study is an extension of \cite{jiang2024artificial}, which uses death data and estimated infection fatality rate (IFR) to back-cast the true daily new COVID-19 infection of all 50 states and Washington DC in the United States. A neural network is then trained to reveal the connection between testing data (testing volume, confirmed infection from testing), population density, and the estimated daily new infection. In this paper, the approach established in \cite{jiang2024artificial} is further extended to include wastewater RNA surveillance data into the training set. The inclusion of wastewater RNA data is crucial because the testing data became increasingly unreliable after the widespread adoption of home antigen tests, beginning in early to mid-2022. The infection fatality data also became less clear when most people in the general population had multiple exposures to the COVID-19 antigen through vaccinations and infections because it is known that for high-risk groups, the severity of COVID-19 highly depends on the time from last vaccination or infection. All these factors make testing data no longer a reliable source for determining community-level infection.

The reliability of testing data and IFR starts to decline in mid-2022, while wastewater viral RNA data collection started in late 2020, and became widely available in 2021. Therefore, there is a short time window where wastewater viral RNA data is available and the true daily infection count can be recovered from death data and IFR. In this paper, we use this time period to establish the connection between daily new infection, wastewater viral RNA concentration, testing data, and other confounding factors such as temperature and precipitation. By training an artificial neural network, we obtain a nonlinear function that maps wastewater viral RNA concentration, testing data, and other data to the daily new infection count. This nonlinear function can be used to estimate the community-level infection of COVID-19 based on new wastewater viral RNA data regardless of whether the testing data is still available.

Compared with \cite{jiang2024artificial}, the main challenge of this study lies in the varying availability of reliable data from different time periods. Early in the pandemic, the data from most states consisted only of testing data. Later in the pandemic, the reliability of both testing data and recovered daily new infection count decreased with time. To address this challenge, we propose using partial derivatives to phase out the dependence on certain neural network inputs. More specifically, when an input variable becomes unreliable, we put the partial derivative with that variable into a loss function to ensure the neural network output is independent of that variable. This approach works well in both the toy example and our neural network training practice. We remark that there have been many published papers attempting to estimate the under-counting of COVID-19 infection \cite{lau2021evaluating, irons2021estimating}, to estimate cases using machine learning tools \cite{xu2022forecasting, zeroual2020deep}, and to connect wastewater surveillance with COVID-19 metrics \cite{li2023wastewater, senaratna2024estimating, hillary2020wastewater}. Compared to these earlier works, we build a comprehensive framework to estimate the true COVID-19 case by using death data and IFR, address the varying reliability of public health data over different period, then use artificial neural network to learn the connection between wastewater viral RNA concentration and the COVID-19 case count. To the best of our knowledge, this has not been reported before. 

Incorporating wastewater RNA data into the training set presents many practical difficulties—numerous methods can be used to measure the viral RNA concentration, and these method can change between time periods even for the same wastewater treatment plant. Data from different entities (e.g., Massachusetts Water Resources Authority, CDC, and Biobot Analytics) also use different metrics. In addition, the wastewater viral RNA data comes from each wastewater treatment plant, while the testing data and recovered daily new cases correspond to statewide totals. After processing  wastewater RNA data, we compared aggregated state-level wastewater RNA data and state-level daily recovered true cases, and selected data from 22 states as our training set. We also include weather data into the training set, as RNA molecules degrade faster in higher temperature, and precipitation can dilute viral RNA in the wastewater system.

The organization of this paper is as follows: \cref{sec:methods} explains the neural network training method, \cref{sec:results} presents our training results, \cref{sec:data_generation} explains the generation of the training set, and \cref{sec:discussion} is the discussion and conclusion. 

\section{Methods}\label{sec:methods}
\subsection{Neural network prediction of true COVID-19 case count}\label{sec:neural_network}
We developed a neural network model under semi-supervised learning to predict the true daily new COVID-19 cases rate, denoted as $C_t$, by using a combination of biological, environmental, and testing-related features. The general form of the prediction function is: 

\begin{equation}\label{eq:basic_model}
C_t = f(W, P, T, \text{Pop}, T_v, T_r, C_c)
\end{equation}

where $W$ is wastewater RNA concentration, $P$ is precipitation, $T$ is daily average temperature, $\text{Pop}$ is population density, $T_v$ is testing volume, $T_r$ is testing rate, and $C_c$ is confirmed case rate. As shown in \cref{eq:basic_model}, the base model incorporates all features.

To account for the declining reliability in testing data, we introduce a time-dependent control parameter $\alpha$:

\begin{equation}\label{eq:alpha_model}
C_t = g(W, P, T, \text{Pop}, T_v, T_r, C_c, \alpha)
\end{equation}

Here, $\alpha$ governs the trustworthiness of both testing data and RNA data over time. This dual mechanism addresses two distinct reliability challenges: (1) Early-phase RNA unreliability (2020), where wastewater RNA monitoring systems were not yet well-calibrated to accurately reflect true COVID-19 case rates; and (2) Late-phase testing unreliability (2021-2022), where widespread use of untracked at-home testing kits introduced in late 2021 deteriorated the reliability of official testing data. As $\alpha$ varies over time, the model adapts its reliance on these features accordingly, as formalized in \cref{eq:alpha_model}.

The model deploys a deep neural network with 6 hidden layers, each containing 512 neurons with swish activation functions. This architecture provides sufficient capacity to capture complex temporal patterns while maintaining computational efficiency. L2 regularization with weight decay $10^{-4}$ is applied to all layers to prevent overfitting.

\subsection{Parameter Categories and Their Influence}
To better understand the influence of each feature, model inputs are categorized as follows:

\paragraph{Biological Indicator: RNA Concentration in Urban Wastewater Systems}
Wastewater RNA concentration ($W$) serves as a valuable biological signal for predicting the true case rate, since SARS-CoV-2 RNA detected in urban wastewater originates only from actively infected individuals. This modeling choice is motivated by prior work \cite{li2023wastewater}, where SARS-CoV-2 RNA concentrations were successfully used to predict weekly new hospital admissions across more than 150 U.S. counties. Unlike testing-derived covariates, $W$ is policy-invariant and provides a direct measure of population-level prevalence. During the early stages of the pandemic (2020), however, wastewater monitoring systems were not yet well-calibrated. To address this, our phase-out mechanism (see below) reduces the model’s reliance on unreliable RNA signals during that period via gradient regularization, while preserving RNA influence in later phases. Empirically, the model learns a strong positive association between $W$ and $C_t$ once RNA monitoring becomes more reliable.

\paragraph{Environmental Factors}
Environmental co-variates influence both the detectability of viral RNA in wastewater and transmission of the virus. 
For example, heavy precipitation can dilute viral concentrations in wastewater, thereby 
reducing the sensitivity of viral detection \cite{saingam2023wastewater}. 
The observed wastewater RNA concentration also depends on water temperature, as RNA molecules degrade faster in warmer conditions. 
However, because wastewater temperature data are unavailable, we instead use average air temperature as a proxy. 
Population density is also positively associated with death rates, since densely populated areas tend to facilitate faster virus spread, resulting in higher case and death counts. 
Taken together, we incorporate these three environmental factors to calibrate the prediction of the true COVID-19 case rate.

\paragraph{Testing-Related Features and Phase-Out Mechanism}
Testing-related features include testing rate ($T_r$), confirmed case rate ($C_c$), and testing volume ($T_v$). A higher testing rate generally increases the probability of detecting existing infections, thereby improving the observed alignment with the true case rate. A higher confirmed case rate ($C_c$) indicates a great epidemic burden in the population, though its reliability depends on the scale of testing \cite{besserve2020assaying}. Larger testing volumes increase the likelihood of capturing more true cases and reduce underdetection bias \cite{lau2021evaluating}. However, these features are subject to policy changes and behavioral shifts, which became less reliable after the introduction of at-home testing kits in late 2021. Thus, their inclusion is modulated by the control parameter $\alpha$.

\subsection{Semi-Supervised Learning Framework}
Our approach utilizes two distinct datasets to implement the phase-out mechanism. The labeled dataset ($\mathcal{D}_L$) contains ground truth COVID-19 rates and is used for supervised learning with mean squared error loss (MSE). Unlabeled samples ($\mathcal{D}_U$) carry time-dependent $\alpha$ values that encode feature reliability and drive the regularization described in the following sections. This design keeps the supervised MSE objective focused on labeled data while using unlabeled data to enforce temporal trust via the $\alpha$-parameterized gradient penalty.

The training process adopts an alternating optimization strategy between supervised learning on labeled data and regularization on unlabeled data, which proves effective for balancing loss components ~\cite{zhai2022deep}. The supervised loss is computed as:
\begin{equation}\label{eq:mse_loss}
\mathcal{L}_{\text{MSE}} = \mathcal{L}_{\text{MSE}}(\mathcal{D}_L)
\end{equation}
and the regularization loss is computed as:
\begin{equation}\label{eq:gp_loss}
\mathcal{L}_{\text{GP}} = \mathcal{L}_{\text{GP}}(\mathcal{D}_U)
\end{equation}
where $\mathcal{L}_{\text{MSE}}$ is computed on labeled training data and $\mathcal{L}_{\text{GP}}$ is the dual gradient penalty computed on unlabeled data that simultaneously penalizes both RNA gradients (during early-phase unreliability when $\alpha < 0$) and testing-related feature gradients (during late-phase unreliability when $\alpha > 0$). This separation ensures that the model learns from reliable ground truth while simultaneously adapting to temporal feature unreliability through $\alpha$-driven gradient suppression, as defined in \cref{eq:mse_loss,eq:gp_loss}. 

To operationalize this temporal trust within the semi-supervised setting, we introduce a time-dependent control parameter $\alpha$ that encodes feature reliability over time. In supervised steps, $\alpha$ is treated as an input context but does not actively participate in the phase-out mechanism; in regularization steps on unlabeled data, $\alpha$ drives the gradient penalty to selectively suppress unreliable features. We next formalize $\alpha(t)$ and show how it parameterizes the phase-out mechanism.

\subsection{Phase-Out Mechanism: Time-Dependent Control Parameter $\alpha$}\label{sec:phase_out}
To translate the temporal feature trust concept into a training strategy, we define a piecewise-linear function $\alpha(t)$ that governs the trustworthiness of both testing data and RNA data over time:
\begin{equation}\label{eq:alpha_function}
\alpha(t) =
\left\{
\begin{array}{ll}
-1 & \quad t \leq \text{December\, 31, 2020} \\[1ex]
0 & \quad \text{December\, 31, 2020} < t \leq \text{August\, 30, 2021} \\[1ex]
\dfrac{t - t_0}{t_1 - t_0} & \quad \text{August\, 30, 2021} < t < \text{March\, 1, 2022} \\[1ex]
1 + \dfrac{t - t_1}{t_1 - t_0} & \quad t \geq \text{March\, 1, 2022}
\end{array}
\right. \,,
\end{equation}

where $t_0 = \text{August\, 30, 2021}$ is the onset of observed testing unreliability and $t_1 = \text{March\, 1, 2022}$ is the point of complete testing unreliability. These time points are selected because household COVID-19 rapid antigen testing became widely adopted in the United States during the transition from the Delta to Omicron waves (August 2021–March 2022), with self-testing among symptomatic individuals rising sharply \cite{rader2022athome}. In addition, the time December 31, 2020 is when wastewater viral RNA data became widely available and reliable. Before this point, many training data points have missing wastewater RNA features. This significantly impacts the training quality regardless of whether we set missing these data points as zero or assign an artificial value to them. Still, that early training data must be included because it is crucial for learning the nonlinear dependency between true case count and testing data. The piecewise-linear function defined in \cref{eq:alpha_function} provides a smooth transition between different reliability phases. 

Function $\alpha(t)$ serves as a time-dependent indicator of feature reliability. The dual regularization behavior is realized at training time through two $\alpha$-driven components:

\begin{itemize}
    \item \textbf{RNA-gradient weight}: Applies heavy penalty weight to RNA gradients when $\alpha<0$ to suppress unavailable or unreliable RNA data during early phases. 
    \item \textbf{Testing-feature weight} $w(\alpha)$: Scales the gradient penalty on $C_c$, $T_v$, and $T_r$ according to reliability, with increasing suppression as $\alpha$ grows.
\end{itemize}

For unlabeled data, $\alpha$ values are randomly sampled from $[-1, 2]$ to provide diverse training scenarios that cover the full spectrum of temporal reliability patterns.

\FloatBarrier
\begin{figure}[H]
    \centering
    \includegraphics[width=0.8\linewidth]{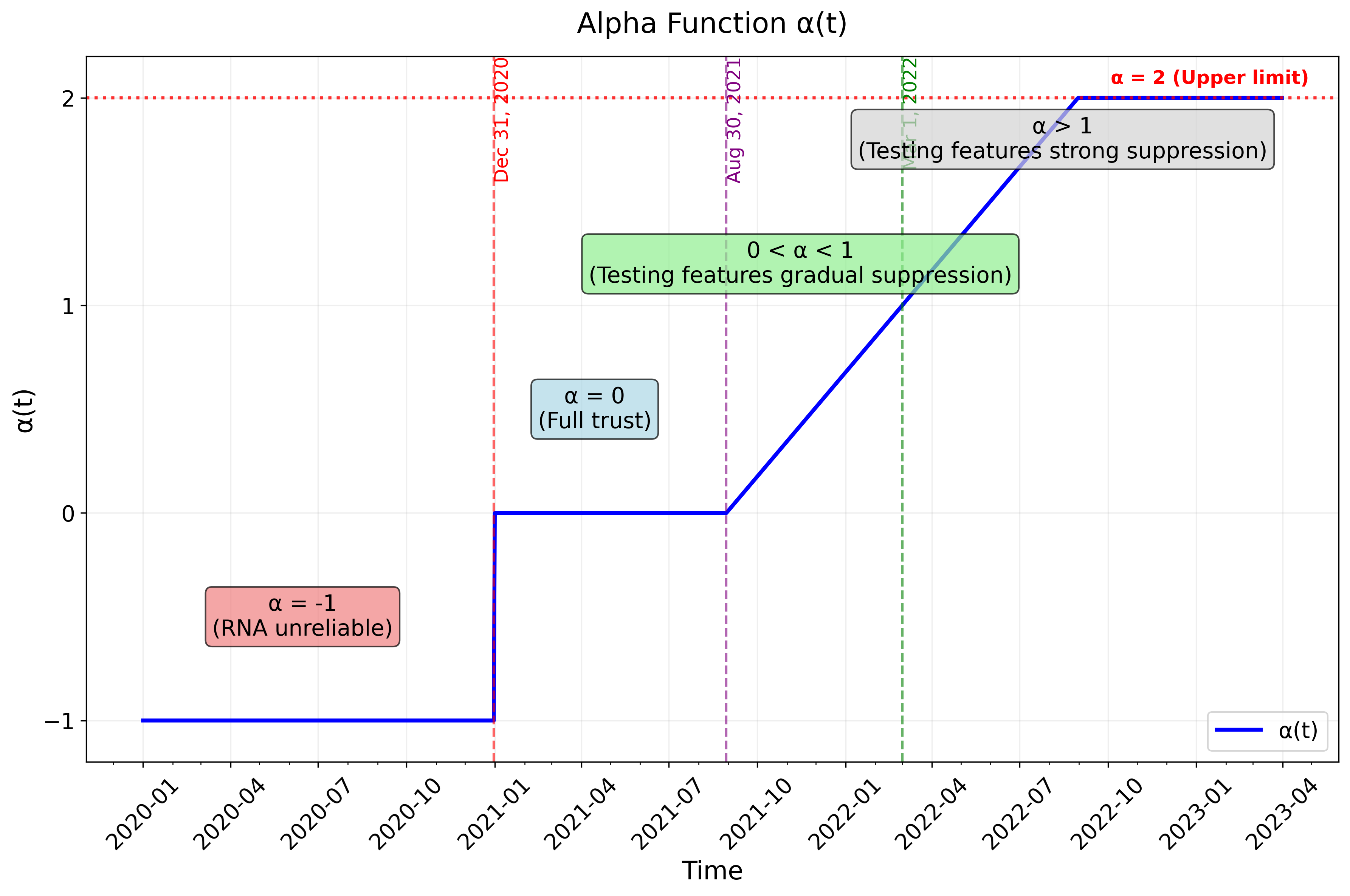}
    \caption{Alpha function $\alpha(t)$.}
    \label{fig:alpha_function}
\end{figure}

Under this setup, when $\alpha = 0$, all features are fully trusted and training gradients flow normally; when $\alpha < 0$, RNA gradients are heavily penalized in the regularization term to effectively suppress their influence; when $0 < \alpha < 1$, the model gradually reduces sensitivity to testing-related features ($C_c$, $T_v$, $T_r$), and when $\alpha \geq 1$, gradients with respect to $C_c$, $T_v$, $T_r$ are strongly suppressed by the gradient penalty.

This mechanism allows the model to \textbf{continuously interpolate between early-phase learning and late-phase robustness}.

\subsection{Gradient Phase-Out: Theory and Implementation}
To provide intuition for the phase-out strategy, we consider a toy function. The toy decay gate $h(\alpha)$ introduced below plays the same conceptual role as the model's $w(\alpha)$ used later: both modulate gradients by reliability encoded in $\alpha$. We define the toy function as:
\begin{equation}\label{eq:toy_function}
y = \sin(x) \cdot h(\alpha) + z^2
\end{equation}
where $x$ represents an unreliable feature, $z$ denotes a stable input, and $h(\alpha)$ is a decay function that modulates the influence of $x$ as $\alpha$ increases over time. We define $h(\alpha)$ as:
\begin{equation}\label{eq:h_function}
h(\alpha)=
\left\{
\begin{aligned}
&1-\dfrac{1}{1+\exp\!\left(-\dfrac{0.3}{1-\alpha}\right)} && \text{if }\alpha<1,\\[0.5ex]
&0 && \text{if }\alpha\ge 1.
\end{aligned}
\right.
\end{equation}
The key observations for the toy function are that as $\alpha \to 1$, $h(\alpha) \to 0$, which causes the $\sin(x)$ term to vanish while the reliable $z^2$ term remains. The gradient $\partial y / \partial x = h(\alpha) \cos(x)$ is smoothly suppressed. It illustrates selective gradient shutoff for unreliable signals. This structure mirrors the behavior we desire in the neural network: preserve stable biological cues while phasing out policy-dependent or late-phase-degraded features, as demonstrated in \cref{eq:toy_function,eq:h_function}.

\FloatBarrier
\begin{figure}[H]
    \centering
    \includegraphics[width=0.8\linewidth]{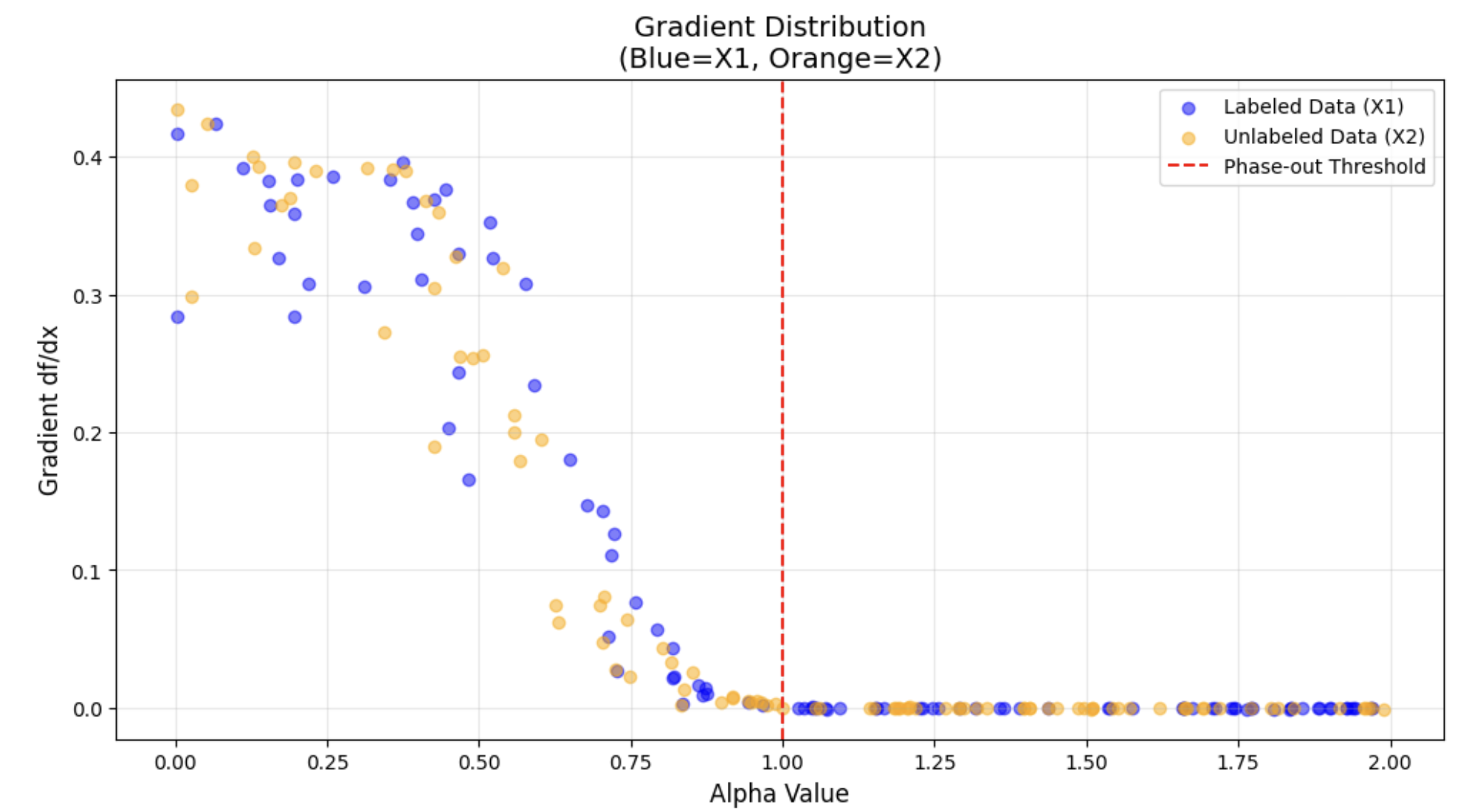}
    \caption{Toy Model: Gradient Distribution of dy/dx with respect to $\alpha$}
    \label{fig:toy_gradient_distribution}
\end{figure}

These two heatmaps respectively display the model's prediction results and the ground truth values. Through comparative analysis, we can verify the model's learning effectiveness at $\alpha$ = 0.5:

\begin{figure}[tbhp]
    \centering
    \subfloat[Prediction heatmap ($\alpha = 0.5$)]{\label{fig:toy_pred}\includegraphics[width=0.48\textwidth]{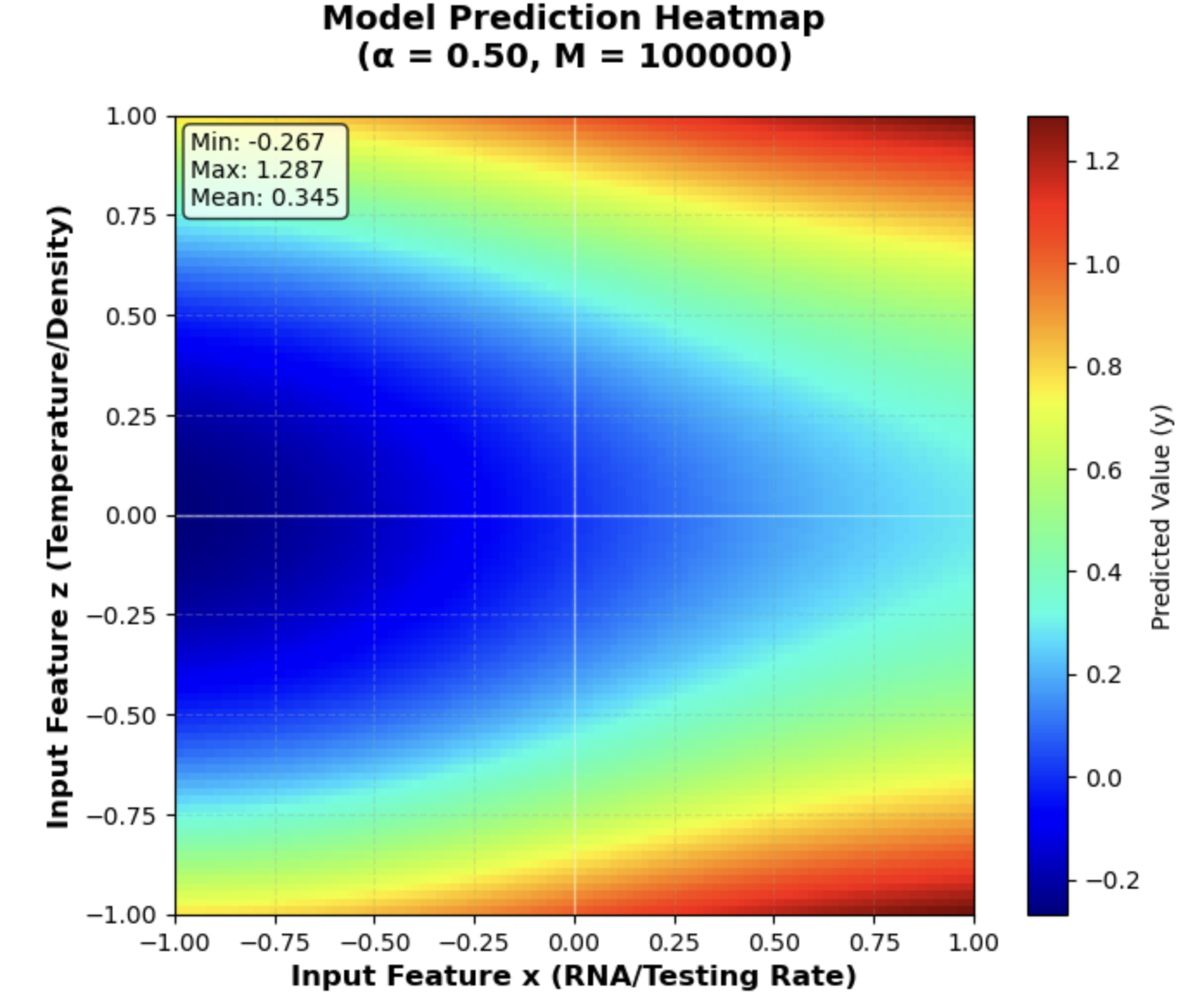}}
    \hfill
    \subfloat[Groundtruth heatmap ($\alpha = 0.5$)]{\label{fig:toy_gt}\includegraphics[width=0.48\textwidth]{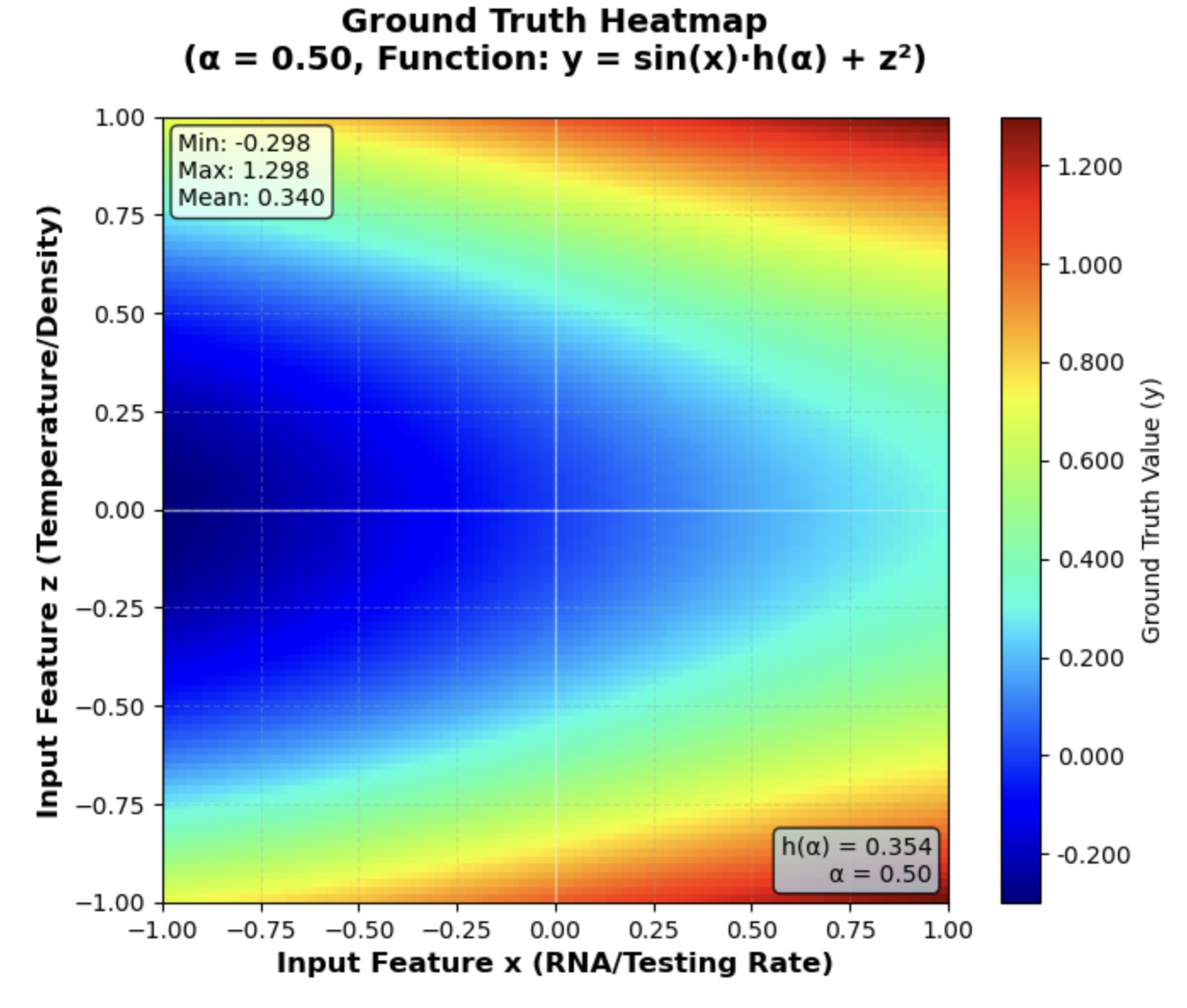}}
    
    \vspace{0.5em} 

    \caption{Toy Model: Prediction vs Groundtruth heatmaps when $\alpha = 0.5$}
    \label{fig:toy_compare}
\end{figure}

Since $h(0.5) \approx 0.5$, the model is in a partially activated state and needs to capture the sinusoidal wave pattern of $\sin(x)$ and the parabolic characteristics of $z^{2}$. 
The visual similarity between the two heatmaps reflects the model's learning quality. 

Building upon the Toy Model's insights, we extend the gradient penalty mechanism to our COVID-19 prediction model. In the Toy Model, the target function $y = \sin(x) \cdot h(\alpha) + z^2$ demonstrates how feature reliability affects model behavior: $x$ represents the testing-related and RNA feature (analogous to testing rate, testing volume, confirmed case rate in our model, and RNA), $z$ represents the reliable feature (analogous to environmental factors like temperature and density), $\sin(x)$ captures the testing feature's pattern, and $h(\alpha)$ acts as a reliability gate that gradually suppresses unreliable testing features as $\alpha$ increases.

In our COVID-19 model, we implement a dual gradient penalty mechanism that handles both RNA and testing feature suppression through $\alpha$-dependent weighting. The mechanism operates in two distinct phases:

\textbf{Testing Feature Suppression:} For testing-related features ($T_r$, $T_v$, $C_c$), we define the weight function $w(\alpha)$ piecewise as:
\[
w(\alpha)=
\begin{cases}
0, & \text{if }\alpha < 0,\\[0.5ex]
\alpha, & \text{if }0 \le \alpha < 1,\\[0.5ex]
1 + \exp(\beta \cdot (\alpha - 1)) - 1, & \text{if }\alpha \ge 1.
\end{cases}
\]
where $\beta > 0$ is the exponential growth rate parameter (optimized by Optuna). This ensures gradual suppression of testing features as their reliability decreases over time.

\textbf{RNA Feature Suppression:} For RNA gradients, we apply a complementary penalty weight:
\[
w_{\text{RNA}}(\alpha) = 
\begin{cases}
100, & \text{if }\alpha < 0,\\[0.5ex]
0, & \text{if }\alpha \ge 0.
\end{cases}
\]
where the large penalty weight (100) strongly suppresses RNA gradients during early-phase unavailability and unreliability.

The complete gradient penalty combines both components, as shown in \cref{eq:gradient_penalty}:
\begin{equation}\label{eq:gradient_penalty}
\mathcal{L}_{\text{GP}} = \mathbb{E}_{x \in \mathcal{D}_U} \left[\, w_{\text{RNA}}\big(\alpha(x)\big) \cdot \left| \tfrac{\partial f}{\partial W} \right| + w\big(\alpha(x)\big) \cdot \frac{\left| \tfrac{\partial f}{\partial T_r} \right| + \left| \tfrac{\partial f}{\partial T_v} \right| + \left| \tfrac{\partial f}{\partial C_c} \right|}{3} \,\right].
\end{equation}

In practice, $\mathcal{L}_{\text{GP}}$ is optimized on unlabeled batches and alternates with the supervised $\mathcal{L}_{\text{MSE}}$ on labeled batches, so that $\alpha$-driven suppression is enforced without contaminating the supervised signal. This realizes the semi-supervised phase-out within the training loop.

This regularization mechanism operates in distinct $\alpha$ ranges: RNA gradients are heavily penalized when $\alpha < 0$ (early-phase RNA unreliability and unavailability), while testing-related gradients are gradually suppressed when $\alpha > 0$ (late-phase testing unreliability). L2 regularization is also applied to all dense layers with weight decay $10^{-4}$ to prevent overfitting.

\subsection{Hyperparameter Optimization}
We employ Bayesian optimization with Optuna \cite{akiba2019optuna} to systematically search the hyperparameter space and identify optimal configurations. The optimization runs 25 trials and uses multi-objective optimization for the gradient penalty version, simultaneously optimizing two objectives.

The first objective is the \textbf{Mean Squared Error (MSE)} on validation data to ensure prediction accuracy, defined in \cref{eq:mse_metric}:
\begin{equation}\label{eq:mse_metric}
\text{MSE} = \frac{1}{n} \sum_{i=1}^{n} (y_i - \hat{y}_i)^2,
\end{equation}
where $y_i$ and $\hat{y}_i$ are the true and predicted values, respectively.

The second objective is the \textbf{Gradient Constraint} to enforce temporal feature trust, which mirrors the dual training regularizer by evaluating both RNA and testing feature suppression across critical phases, as defined in \cref{eq:gradient_constraint}:
\begin{align}\label{eq:gradient_constraint}
\text{Gradient Constraint} &= \mathbb{E}_{x \in \mathcal{D}_U:\ \alpha(x) < 0 \text{ or } \alpha(x) > 1} \left[\, w_{\text{RNA}}\big(\alpha(x)\big) \cdot \left| \tfrac{\partial f}{\partial W} \right| \right. \nonumber \\
&\quad \left. + w\big(\alpha(x)\big) \cdot \frac{\left| \tfrac{\partial f}{\partial T_r} \right| + \left| \tfrac{\partial f}{\partial T_v} \right| + \left| \tfrac{\partial f}{\partial C_c} \right|}{3} \,\right].
\end{align}

This metric captures the dual suppression mechanism: RNA gradients are penalized when $\alpha < 0$ (early-phase RNA unreliability), while testing-derived feature gradients are penalized when $\alpha > 1$ (late-phase testing unreliability), ensuring that sensitivity to unreliable features is effectively suppressed across both critical phases.

\FloatBarrier
The hyperparameter search space for optimizing these objectives is summarized in \cref{tab:hyperparams}.

\begin{table}[tbhp]
\centering
\caption{Hyperparameter Search Space for Optuna Optimization}
\label{tab:hyperparams}
\begin{tabular}{lll}
\toprule
\textbf{Parameter} & \textbf{Range/Values} & \textbf{Description} \\
\midrule
Learning Rate $\eta_1$ & $[10^{-5}, 10^{-3}]$ & Primary learning rate \\
Learning Rate $\eta_2$ & $[10^{-6}, 10^{-4}]$ & Secondary learning rate \\
Epochs & $[10, 30]$ & Training iterations \\
Batch Size & $\{64, 128, 256\}$ & Mini-batch size \\
Hidden Units & $\{96, 128, 256, 512\}$ & Network width \\
Decay Steps & $\{500, 800, 1000, 2000\}$ & LR decay frequency \\
Decay Rate & $[0.8, 0.99]$ & LR decay factor \\
$\lambda_{\text{reg}}$ & $[0.1, 2.0]$ & Training regularization strength \\
$\beta$ & $[0.5, 5.0]$ & Gradient weight slope for $w(\alpha)$ \\
\bottomrule
\end{tabular}
\end{table}

The multi-objective optimization returns Pareto-optimal solutions that balance prediction accuracy against temporal feature trust.

\section{Results and Evaluation}\label{sec:results}
\subsection{Experimental Setup}

\paragraph{Data Construction and Preprocessing.} We construct 13,000 state-day level \newline records by merging wastewater RNA ($W$), precipitation ($P$), temperature ($T$), population density ($Pop$), confirmed case rate ($C_c$), testing volume ($T_v$), and testing rate ($T_r$) from 2020-2023 for 22 states. Data before March 1, 2020 is excluded due to extensive missing values in this early period. Missing entries are set to zero. Features are scaled consistently as follows: log-transform for $W$ and Pop-derived density, square-root or cube-root transforms for $C_c$ and $T_v$, and linear scaling for $P$, $T$, and $T_r$. With only 6,000 RNA observations out of 13,000 total records, missing RNA values are imputed using temporal interpolation from the nearest available measurement within the same state.

\paragraph{Training Configuration.} In the experimental implementation, $\alpha(t)$ follows the piecewise-linear schedule defined in Section 2.2 for labeled data, while for unlabeled data, $\alpha$ values are randomly sampled from $[-1, 2]$ to provide diverse training scenarios for the gradient penalty mechanism. The validation set consists of the last contiguous 2,000 labeled samples from the training time series sequence in order to maintain temporal order and evaluate the model's ability to generalize to future time periods. We evaluate two regimes: (i) semi-supervised (hybrid) and (ii) supervised baseline, following the training procedure described in Methods. The network uses 6 dense layers with 512 hidden units (swish, L2=$10^{-4}$); Adam with two learning rates ($\eta_1=8\times10^{-5}$, $\eta_2=8\times10^{-5}$) and gradient clipping at 0.5. In the hybrid regime, $\lambda_{reg}$ warms from 0.1 to 0.5 over the first 10 epochs and then stays at 0.5. For reporting, we present MSE and MAE of the training dataset.

\subsection{Results}
We evaluate the gradient penalty mechanism across all 22 states in order to assess its generalization capabilities. 

Massachusetts (MA) serves as our primary evaluation target due to its comprehensive and high quality data coverage, with continuous observations from 2020 to now without any temporal interruptions. The state provides 850 consecutive data points in our training set. This completeness enables evaluation of the temporal phase-out mechanism across all $\alpha$ regimes. 

Other states experience significant data gaps between late 2020 and late 2021 due to missing observations of wastewater viral RNA data. These gaps limit their ability to demonstrate the full temporal evolution of feature reliability. 

The following analysis first presents aggregate performance metrics (MSE, MAE) across all 22 states to compare the gradient mechanism against the baseline. Subsequently, we examine detailed results for three representative states: Massachusetts (comprehensive coverage), Arizona and New York (both showing strong performance improvements with the proposed mechanism).

\subsubsection{Gradient Distribution Analysis}
\begin{figure}[tbhp]
    \centering
    \includegraphics[width=0.8\textwidth]{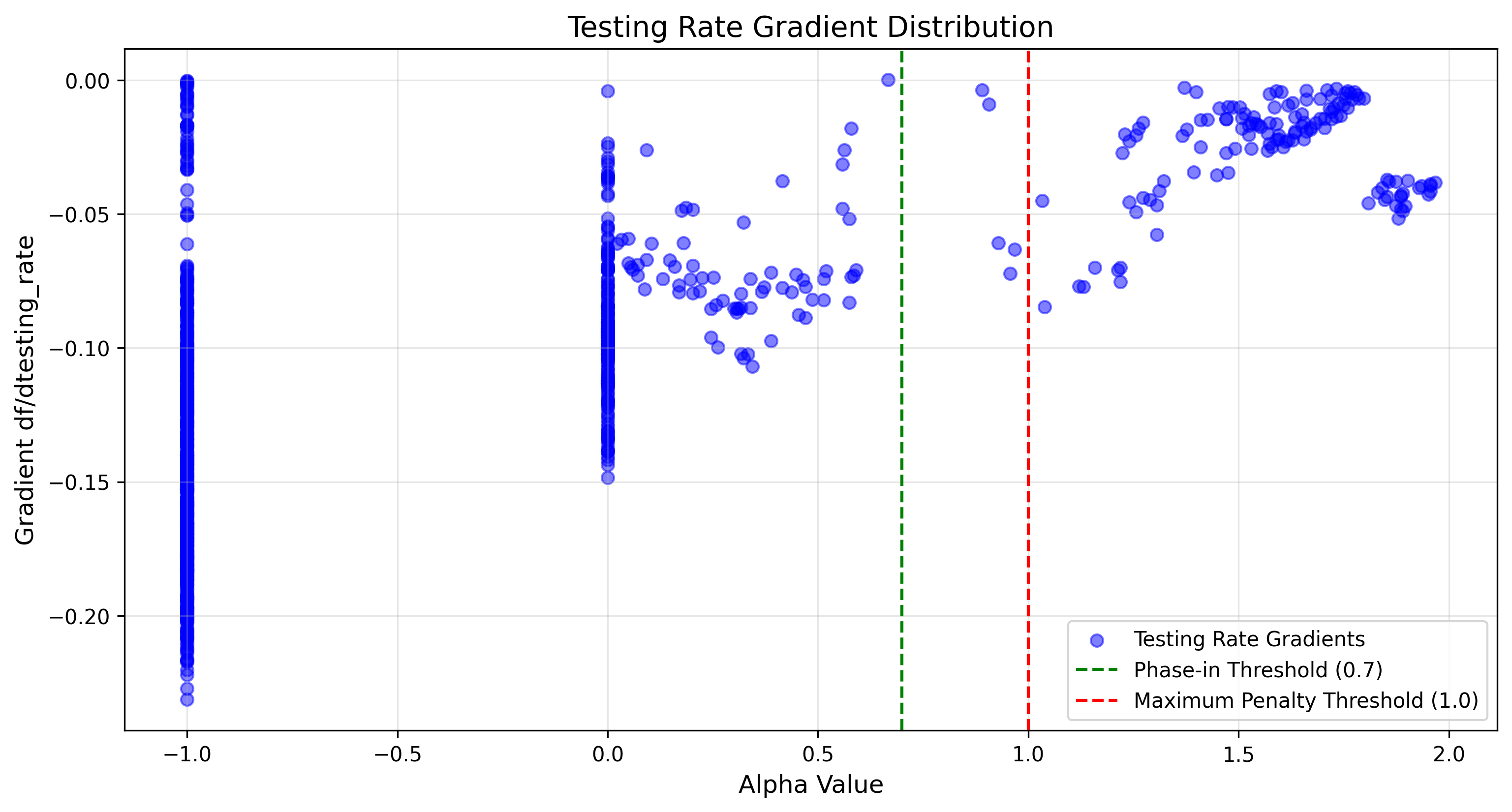}
    \caption{Gradient distribution analysis showing the suppression of testing-related feature gradients as $\alpha$ increases. The mechanism effectively reduces gradient magnitudes when $\alpha > 1$.}
    \label{fig:gradient_distribution}
\end{figure}

To validate the effectiveness of our phase-out mechanism, we analyze the gradient distributions of the model's predictions with respect to testing-related features ($C_c$, $T_v$, $T_r$) across different $\alpha$ values. \cref{fig:gradient_distribution} shows the gradient magnitude distributions computed on a random sample of 1,000 training data points from all 22 states. As shown, when $\alpha > 1$, the gradients are substantially suppressed toward zero, demonstrating that the mechanism effectively reduces testing-related gradients as $\alpha$ increases beyond 1. Importantly, the gradients do not converge to exactly zero, as complete suppression would eliminate the model's ability to learn patterns from testing data. Although testing data became less accurate after August 2021, it still serves as a valuable feature for predicting the true case rate. Perfect convergence to zero would result in reduced prediction accuracy in terms of MSE and MAE. This represents a carefully calibrated trade-off between reducing reliance on unreliable features while preserving their residual predictive value.

\FloatBarrier
\subsubsection{Aggregate Performance Across 22 States}
\cref{tab:aggregate_performance} presents the mean performance metrics across all 22 states, to compare the baseline supervised approach with our gradient phase-out mechanism.

\begin{table}[tbhp]
\centering
\caption{Aggregate Performance Comparison Across 22 States (Mean Values)}
\label{tab:aggregate_performance}
\begin{tabular}{l|ccc}
\toprule
\textbf{Metric} & \textbf{Baseline} & \textbf{Mechanism} & \textbf{Improvement} \\
\midrule
MSE & 0.76 & 0.66 & +12.8\% \\
MAE & 0.64 & 0.54 & +15.4\% \\
\bottomrule
\end{tabular}
\end{table}
The gradient phase-out mechanism demonstrates consistent improvements across all metrics when averaged over 22 states. The mechanism achieves a 12.8\% reduction in MSE and 15.4\% improvement in MAE compared to the baseline.

\subsubsection{State-Specific Performance Analysis}
\cref{tab:state_specific_performance} provides detailed results for three representative states: Massachusetts (comprehensive temporal coverage across all $\alpha$ phases), Arizona and New York (both show strong performance improvements despite data gaps).

\begin{table}[tbhp]
\centering
\caption{State-Specific Performance: Massachusetts, Arizona, and New York}
\label{tab:state_specific_performance}
\resizebox{\textwidth}{!}{ 
\begin{tabular}{l|cc|cc|cc}
\toprule
& \multicolumn{2}{c|}{\textbf{Massachusetts}} & \multicolumn{2}{c|}{\textbf{Arizona}} & \multicolumn{2}{c}{\textbf{New York}} \\
\textbf{Metric} & \textbf{Baseline} & \textbf{Mechanism} & \textbf{Baseline} & \textbf{Mechanism} & \textbf{Baseline} & \textbf{Mechanism} \\
\midrule
MSE & 1.27 & 1.26 & 0.66 & 0.55 & 0.61 & 0.50 \\
MAE & 0.91 & 0.85 & 0.52 & 0.43 & 0.52 & 0.42 \\
\bottomrule
\end{tabular}
}
\end{table}

Massachusetts shows moderate improvements with 5.7\% MAE reduction. Arizona demonstrates strong performance gains with 16.3\% MSE reduction and 18.7\% MAE improvement. New York achieves 17.4\% MSE reduction and 18.6\% MAE improvement. These results indicate the mechanism's effectiveness across different state characteristics. 

\subsubsection{State-wise Time Series Predictions}
\begin{figure}[tbhp]
    \centering
    \includegraphics[width=0.8\linewidth]{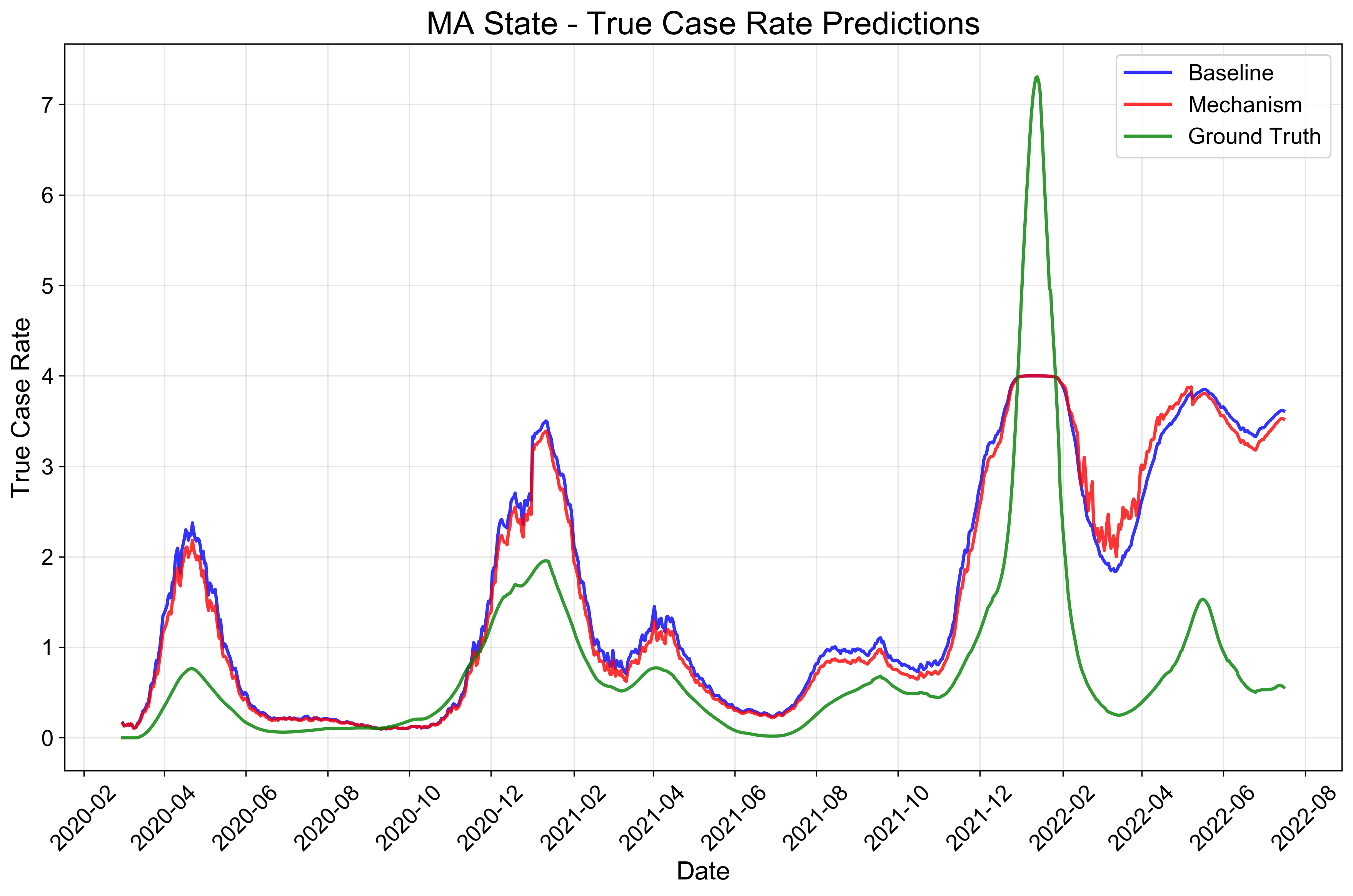}
    \caption{Massachusetts predictions comparison}
    \label{fig:MA_predictions}
\end{figure}

\begin{figure}[tbhp]
        \centering
        \includegraphics[width=0.8\linewidth]{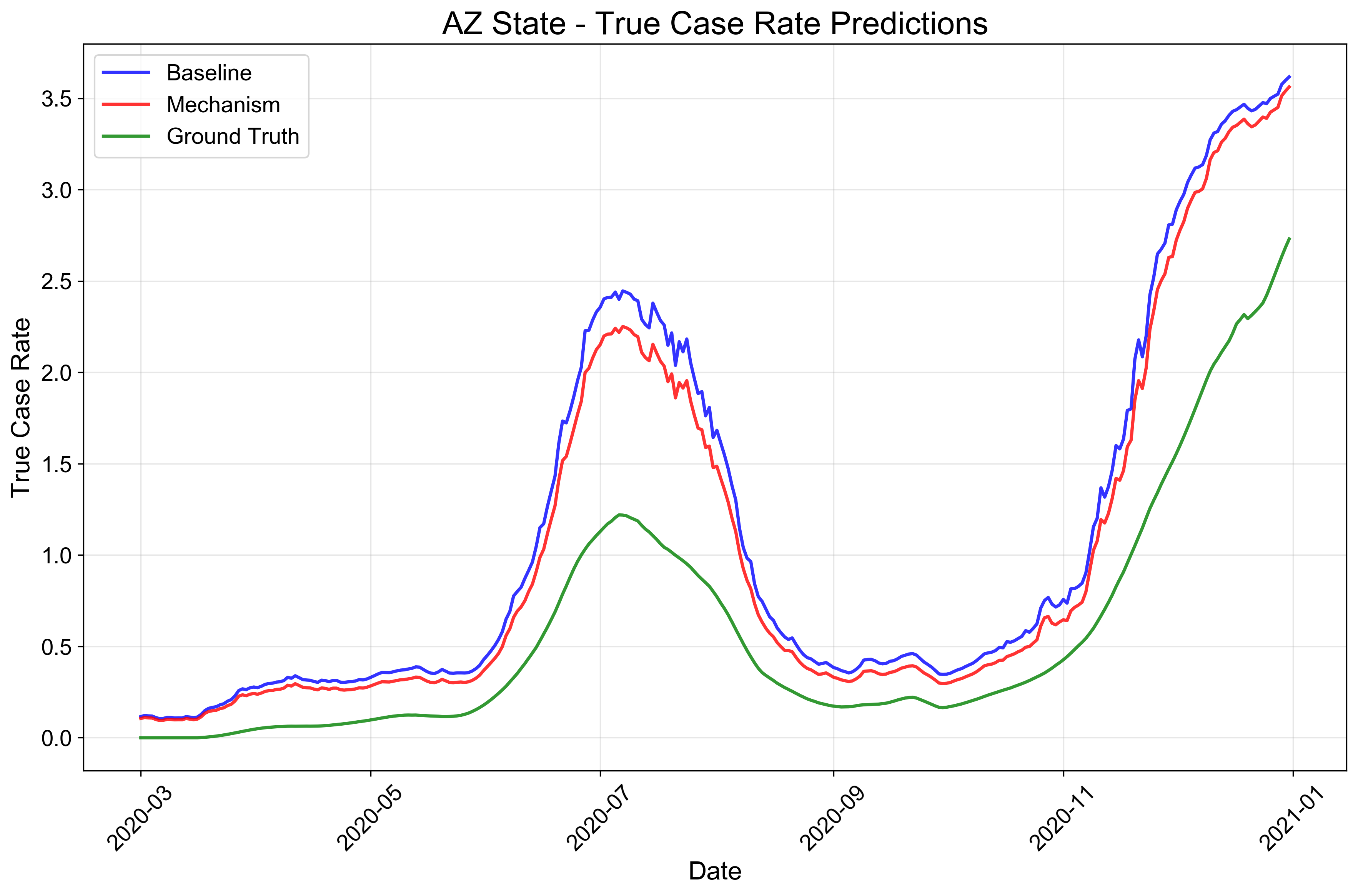}
        \caption{AZ predictions comparison}
        \label{fig:AZ_predictions}
    \end{figure}

\begin{figure}[H]
        \centering
        \includegraphics[width=0.8\linewidth]{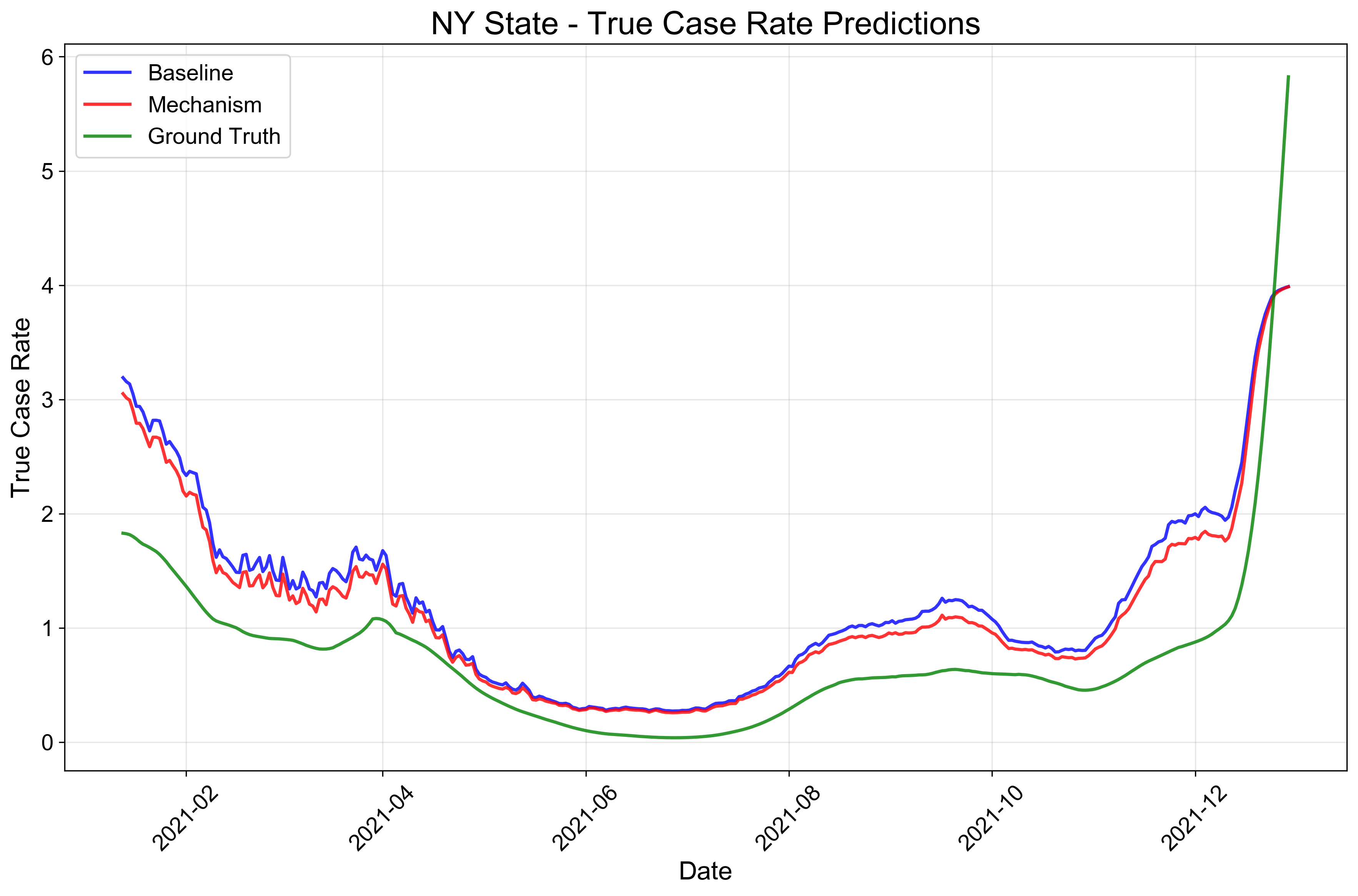}
        \caption{NY predictions comparison}
        \label{fig:NY_predictions}
    \end{figure}

\cref{fig:MA_predictions,fig:AZ_predictions,fig:NY_predictions} presents time series predictions for Massachusetts, Arizona, and New York, comparing the baseline and gradient phase-out approaches against ground truth values.

We remark that the "ground truth" is actually the true cases we recovered from the death data and IFR, not the reported cases. In fact, it is very possible that our method in \cite{jiang2024artificial} also under-counted the true cases to some degree, because many early COVID-19 death were not properly linked to COVID-19 when the testing capacity was very limited. In other words, the ``Ground Truth" in Figure \ref{fig:MA_predictions}, Figure \ref{fig:NY_predictions}, and Figure \ref{fig:AZ_predictions} should be higher. Therefore, the gap between the predicted value and the ground truth is in fact a combination of the prediction error and the correction of the recovered true cased based on the wastewater viral RNA data. 

\subsubsection{Two-Dimensional Heatmap Analysis}
\begin{figure}[H]
     \centering
    \includegraphics[width=1\linewidth]{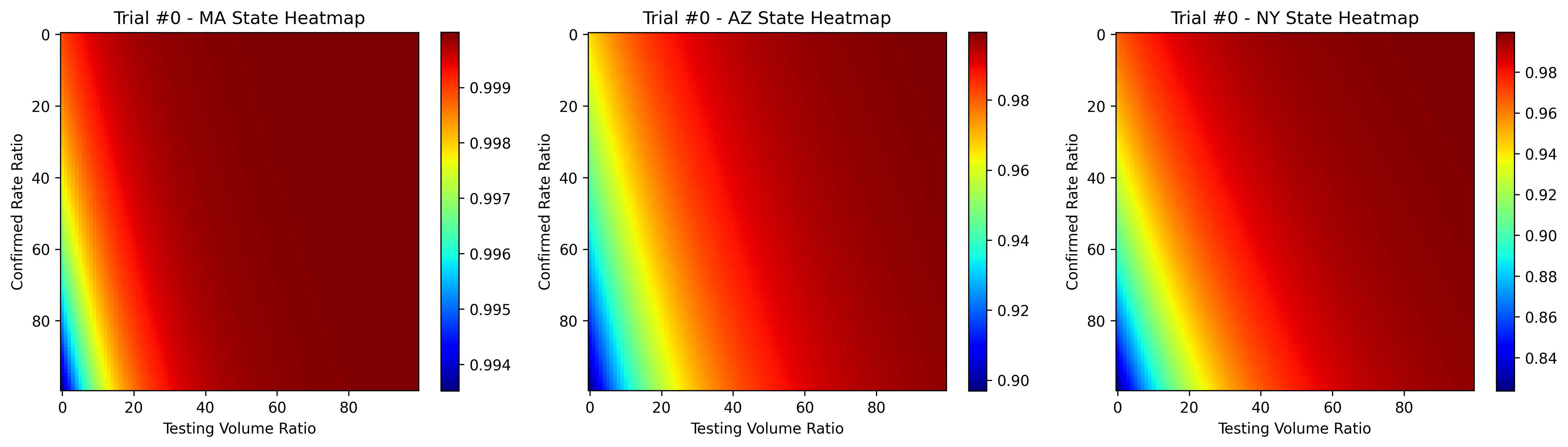}
    \caption{Two-Dimensional Heatmap for Gradient Mechanism}
    \label{fig:mechanism_heatmap}
\end{figure}

\begin{figure}[H]
    \centering
    \includegraphics[width=1\linewidth]{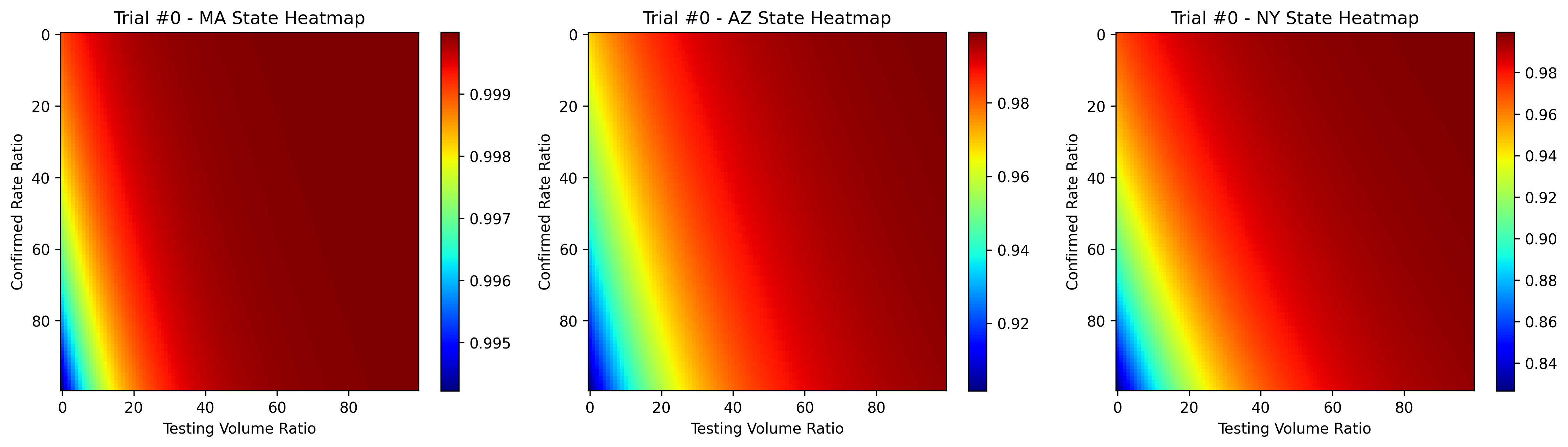}
    \caption{Two-Dimensional Heatmap for Baseline}
    \label{fig:baseline_heatmap}
\end{figure}

\cref{fig:mechanism_heatmap,fig:baseline_heatmap} \newline presents heatmaps that visualize the relationship between testing volume ratio, confirmed rate ratio, and predicted true case counts for Massachusetts, Arizona, and New York. The heatmaps are generated using test data from January 15, 2022, which was selected as a representative date during the alpha growth phase (when $\alpha \approx 0.4$). This choice allows us to evaluate the model's behavior during the critical transition period when feature reliability is evolving. 

\subsubsection{Best Trial Hyperparameters}
\cref{tab:best_hyperparameters} presents the optimal hyperparameters found by Optuna for both training regimes, which highlights the key differences in configuration that contribute to the performance improvements.

\begin{table}[H]
\centering
\caption{Best Trial Hyperparameters: Baseline vs. Gradient Phase-Out}
\label{tab:best_hyperparameters}
\begin{tabular}{l|cc}
\toprule
\textbf{Parameter} & \textbf{Baseline} & \textbf{Mechanism} \\
\midrule
Learning Rate $\eta_1$ & 1.95e-04 & 1.95e-04 \\
Learning Rate $\eta_2$ & N/A & 3.18e-06 \\
Hidden Units & 512 & 512 \\
Batch Size & 256 & 256 \\
Epochs & 28 & 28 \\
$\lambda_{\text{reg}}$ & N/A & 0.60 \\
$\beta$ & N/A & 1.37 \\
\bottomrule
\end{tabular}
\end{table}

\section{Generation of training set}\label{sec:data_generation}
\subsection{True case count}
 

The true daily COVID-19 case count $C_t$ plays a crucial role in our model because it provides the groundwork on which our future predictions are calibrated to and validated on. Note that $C_t$ is very different from the confirmed case count, as many COVID-19 cases remain undetected by the public health agencies. Here, we roughly follow \cite{jiang2024artificial} to recover the true daily case count from the death data and the infection fatality ratio (IFR). As per \cite{jiang2024artificial}, the general structure for our recovered true case count follows:
$$
\mbox{daily new case} \: * \space \mbox{delay} \times \mbox{IFR} = \mbox{daily new death},
$$
where $*$ is convolution. For the sake of completeness of the paper, we review the method used in \cite{jiang2024artificial} along with extensions of our own to expand the training set.

\subsubsection{Backcasting}

We mainly rely on the backcasting method to recover the true daily COVID-19 case count since the IFR for COVID-19, particularly before the introduction of vaccines and novel treatment methods, has been intensively studied. In addition, compared to the confirmed case count, the confirmed death count of COVID-19 is relatively reliable. Therefore, we have:
$$
D_n = \sum_{i = 0}^n I_i \delta_{n-i} \xi_i \,,
$$
where $D_n$ is the death count on day $n$, $I_i$ is the true new infection count on day $i$, $\delta_r$ is the distribution of time delay from a new infection to a confirmed death such that $\delta_r = \mathbb{P}[\mbox{deaths occurred after } r \mbox{ days}]$, and $\xi_i$ is the IFR on day $i$. Since $I_i$ is unknown, this becomes a typical de-convolution. As introduced in \cite{jiang2024artificial}, we first use a time series of confirmed cases and confirmed deaths to estimate the delay distribution $\delta$. Then we solve the deconvolution problem using the confirmed death data and the estimated time series of the IFR. 

Note that the deconvolution is inherently unstable, as small fluctuations in the death count can lead to significant amplifications in the infection count. To address this, we apply the regularization technique introduced in \cite{jiang2024artificial, miller2022statistical} by penalizing large second order derivatives and fourth order derivatives. With regularization, the deconvolution problem becomes a least square problem, whose solution is the true daily new infection count.

\subsubsection{Role of the Infection Fatality Ratio (IFR)}

As is in \cite{jiang2024artificial}, the IFR measures the proportion of COVID-19 infections that leads to death. This value varies based on several factors, including case age distribution, variants of COVID-19, and vaccination rates.

\begin{figure}
    \centering
    \includegraphics[width=1\linewidth]{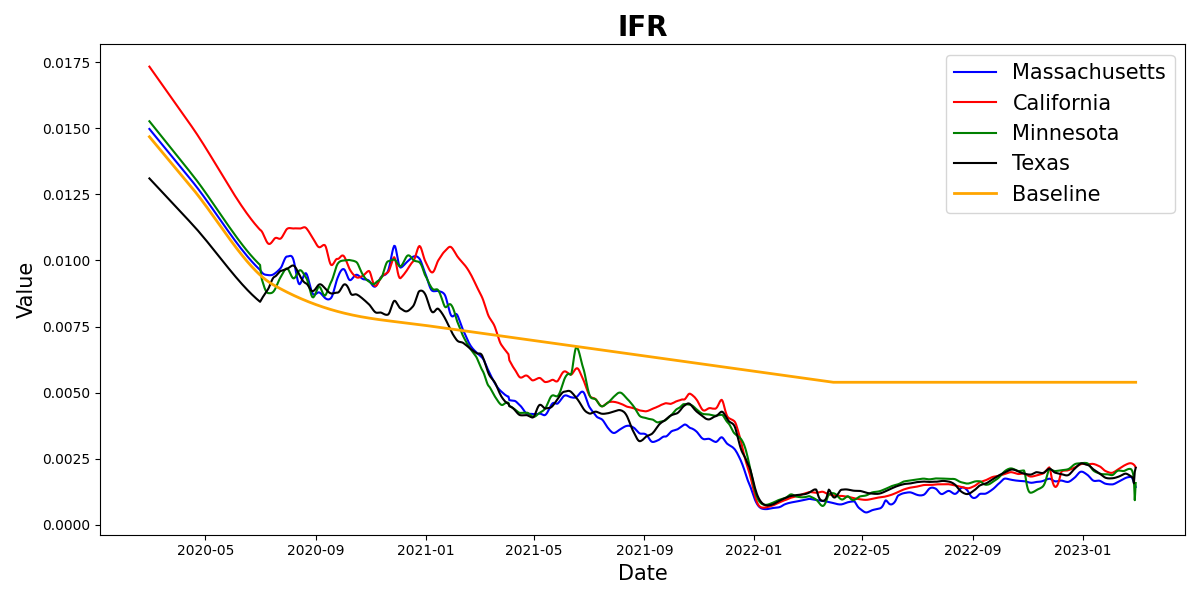}
    \caption{IFR data from 29 Feb. 2020 to 1 Mar. 2023 for the states MA, CA, MN, TX, and the baseline IFR}
    \label{fig:total_IFR}
\end{figure}

In all, the IFR at time $t$ can be expressed as:
\setlength{\abovedisplayskip}{4pt}
\setlength{\belowdisplayskip}{4pt}
\[
\mathrm{IFR}(t)=\mathrm{IFR}_b(t)\times\mathrm{IFR}_T(t)\times
\mathrm{IFR}_A(t)\times\mathrm{IFR}_V(t)\times\mathrm{IFR}_O(t)
\]
where IFR$_b$ is the baseline IFR, IFR$_T$ is the decreasing IFR due to improved treatment, IFR$_A$ is the changes in IFR from different age distributions, IFR$_V$ is the changes to IFR from vaccination count, and IFR$_O$ is the changes to IFR due to variations in Omicron. As demonstrated in \cref{fig:total_IFR}, IFR is generally decreasing until 2022, which is what we would expect as vaccination effectiveness and accessibility increase. Estimation method of $\mbox{IFR}(t)$ is introduced in \cite{jiang2024artificial}, where the baseline IFR, age dependency of IFR, and the time dependent decrease of IFR are estimated based on \cite{sorensen2022variation, barber2022estimating}. Age distribution of cases, vaccination data, and outcome after vaccination comes from the CDC database \cite{CDCcaserate,CDCvacstat,CDCvacstat}. The change of variants comes from CDC database \cite{CDCvariant}.

When extending the IFR for each state into 2023, we found that it became increasingly difficult to estimate. Compared to 2021, health departments in 2022 and 2023 reduced the frequency of death reports from daily to weekly or even biweekly. Subsequently, daily death rates became less accurate, and so did the daily baseline IFR. In addition, after 2023 the majority of the population has more than one COVID-19 infections, which makes IFR estimation increasingly harder. In 2022 and onward, the data we collected has strange fluctuations that did not align between states. To minimize these inconsistencies, we held the baseline IFR of each state constant after 31 Mar. 2022. This trend is reflected in \cref{fig:total_IFR}. As a general trend, the baseline IFR follows what we would expect: as treatment and awareness improved, the fatality rate dropped. Even after March 2022, though, total IFR should not be a constant (even if baseline IFR is). Thus IFR$_T$, IFR$_A$, IFR$_V$, and IFR$_O$ continue to influence the IFR rate.

\begin{figure}[H]
    \centering
    \includegraphics[width=1\linewidth]{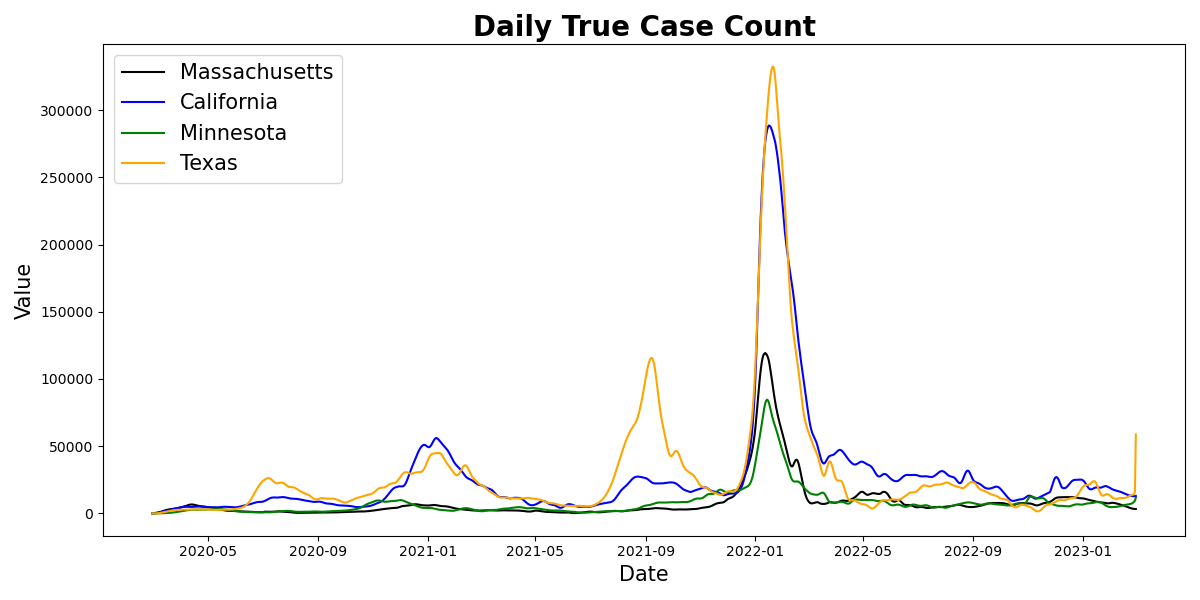}
    \caption{Recovered daily true case count from 29 Feb. 2020 to 1 Mar. 2023 for the states MA, CA, MN, and TX}
    \label{fig:true_case_count}
\end{figure}

\subsection{Testing data}


COVID-19 testing data is used in this study as an input for understanding the relationship between reported case counts and the true scale of infections. Our analysis leverages the Coronavirus Resource Center of Johns Hopkins University (JHU) database, which provides comprehensive daily testing data across the United States. The testing data we use includes the daily testing volume and daily confirmed case count for each state plus Washington DC. The use of testing data is similar to that in \cite{jiang2024artificial}. 

One important remark is that the testing data becomes increasingly unreliable after March 2022 because more people chose to test COVID-19 with home antigen test instead of PCR tests. Therefore, many cases are not reported to the public health department any more. That is why we need to gradually phase out the dependency on testing data in our training set. 

\subsection{Wastewater RNA concentration}


The selection of the training data was aimed at ensuring a high level of data integrity and relevance for modeling true COVID-19 case dynamics based on viral wastewater RNA concentration. The data utilized in this study were collected from multiple sources, including the Massachusetts Water Resources Authority (MWRA) \cite{masscovid}, Biobot, the Biobot Inc, and the Centers for Disease Control and Prevention (CDC) \cite{CDCwastewater}. This comprehensive dataset encompasses RNA measurements collected from various wastewater treatment facilities across the United States. It provides a unique vantage point on the spread of COVID-19, independent of clinical testing rates.


Initial data cleaning involved the normalization of data formats and the handling of incomplete records. Specifically, inconsistencies in date formats across datasets were rectified by converting all dates to a standardized format. For missing values in critical fields such as the wastewater viral RNA, we opted to assign a value of zero. Since training set with missing viral RNA data concentrates in data from 2020, we then suppress the dependence between the recovered true case and the viral RNA in this data. This method also prevents the introduction of potential biases that might arise from imputing these values based on neighboring or artificial data points.

The initial dataset was comprised of RNA concentration readings at the county level, or more precisely, the county name of the wastewater treatment plant. To align this data with the state-level analyses and enhance statistical robustness, we aggregated these county-level data into state-level estimates. The aggregation was performed by calculating the weighted average of RNA concentrations of counties within each state. Since RNA concentration essentially measures the contributions of each individual to a community's total viral RNA data, we can measure a state's wastewater RNA concentrations by proportionally cumulating the county data based on population served by each wastewater treatment facility. So this aggregation takes into account varying sizes of population density to prevent skewed results favoring more densely populated areas. Overall, our state RNA measurements reflect a population-weighted per capita rate, which is more representative of the state’s overall viral load.

Each county's RNA data was mapped to its corresponding state based on the geographical location of the wastewater treatment facilities. Using the FIPS codes of each wastewater facility, we mapped them to their county, and then subsequently state. Demographic statistics of each county were then collected to be used in calculating state-level RNA concentrations. Each facility within any given state was used as part of the state-wide average.




\subsubsection{Comparative Analysis and Data Set Selection}
After manipulating the RNA data, we compared each state's CDC and Biobot data to the true cases we collected previously. This was done through graphical visualizations of the true daily new case data and either CDC or Biobot across the same time periods. We then selected periods where the RNA data followed similar trends to the true case data. Specifically for Massachusetts, we found that the MWRA data fits well with the trends of our true cases. Our resulting RNA data set is a collection of CDC, Biobot, and MWRA data in selected time periods for specific states that match trends with our true case data of the same time period in the same state.

\begin{figure}[H]
    \centering
    \subfloat[Florida's CDC data compared to the true case count]{\label{fig:florida_cdc_vs_true_case}\includegraphics[width=0.5\textwidth]{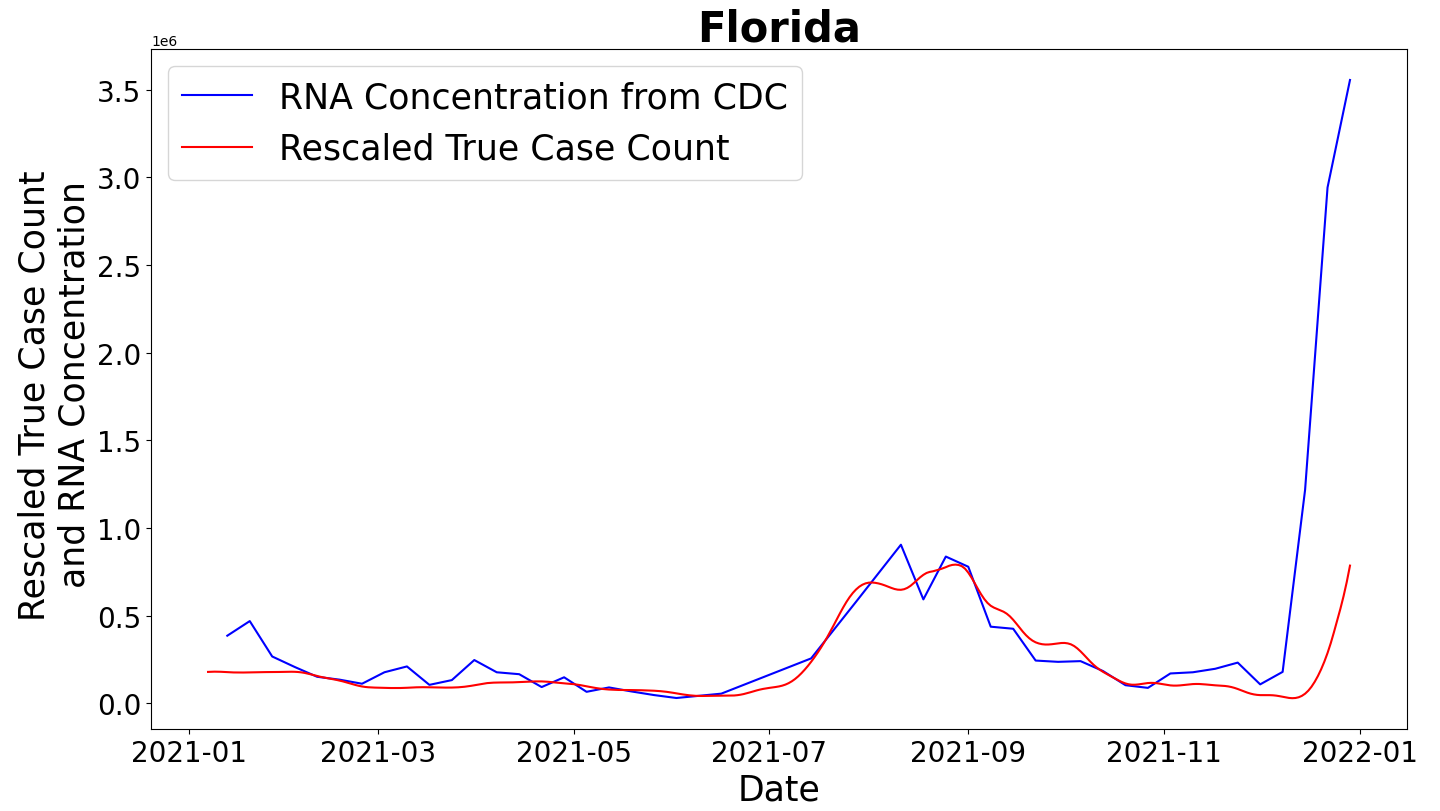}}
    \hfill
    \subfloat[Michigan's CDC data compared to the true case count]{\label{fig:michigan_cdc_vs_true_case}\includegraphics[width=0.5\textwidth]{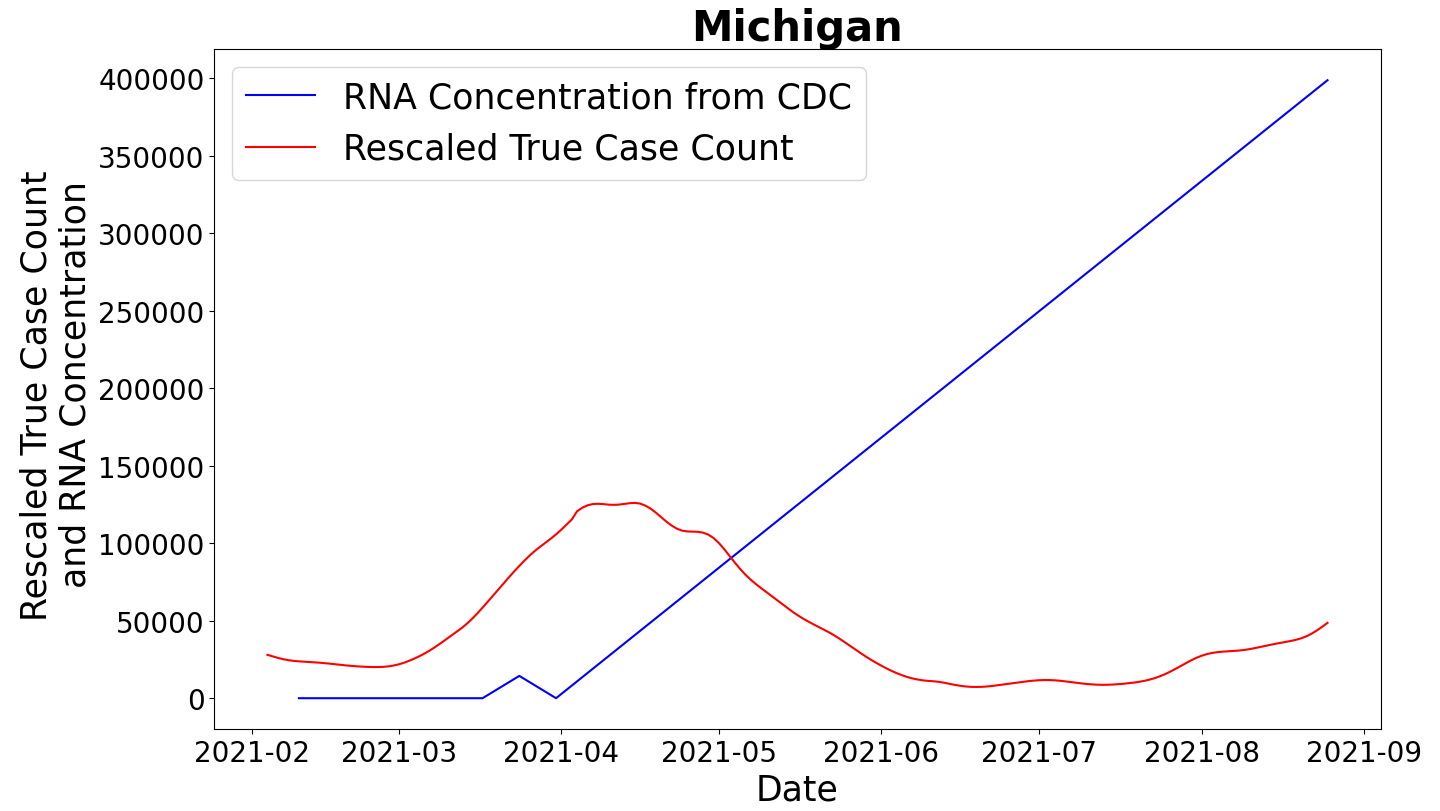}}
    \caption{Comparison of CDC wastewater RNA data with true case counts for Florida and Michigan. Note that true case counts were rescaled by 8, though we are only concerned with the similarities/differences between the true case count and RNA concentration trends, so exact values do not matter.}
    \label{fig:florida_michigan_comparison}
\end{figure}

In general, CDC data was the least consistent with our true case calculations, Biobot in between, and MWRA data the most consistent. Note that, even with its inconsistencies, CDC data was still used when its graphs with true case count aligned. For example, in \cref{fig:florida_cdc_vs_true_case}, Florida's CDC data and true case count are fairly consistent, allowing us to use it in our training. Conversely, in \cref{fig:michigan_cdc_vs_true_case}, Michigan's CDC data does not match well with the true case count, and consequently is not used in our training set.

We found confirmed cases to be valuable in the early stages of COVID-19, since testing was almost entirely occurring in medical institutions, and thus recorded as official COVID-19 cases. When rapid antigen self-tests became available and people started testing at home more frequently, confirmed cases were rendered inaccurate. As a result, the recovered daily new case count becomes less reliable as well. Thus, we must phase out confirmed cases as self-testing becomes prevalent. This renders RNA concentrations as the predominant data source. Conversely, we also phase out RNA concentrations earlier in COVID-19's transmission, making confirmed cases the focus during that time period. In all, this ensures the training set data is as accurate as possible, dynamically emphasizing data sources when they become the most accurate.

\setlength{\intextsep}{6pt}
\setlength{\intextsep}{12pt}
\subsection{Weather data}

Weather data are important in our dataset because they allow us to account for RNA data decay. We found that fluctuations in temperature and precipitation affect the degradation of viral RNA in wastewater. Under harsher weather conditions and higher temperatures, RNA decays more rapidly. Therefore, collecting weather data is vital for maintaining the fidelity of our analysis.

Our weather data were collected from the National Oceanic and Atmospheric Administration (NOAA) website. Using the Climate Data Online Search tool, we collected records from 1 January 2020 to 21 June 2024 for weather stations in each U.S. state. When selecting the number and locations of stations for each state, we considered the state population sizes. We selected the two to four most populous counties in each state. The weather stations in these counties were identified through NOAA, and their Daily Summaries datasets were downloaded.

Each dataset contained various hourly  measurements, including dry-bulb temperature and precipitation. If any hourly measurements were missing during a day, we replaced them with zeros. We then averaged these values over each day to obtain the daily mean temperature and precipitation. Consequently, we obtained daily average temperature and precipitation values for two to four areas in each state.

\section{Discussion and Conclusion}\label{sec:discussion}
\subsection{Performance Analysis and Key Findings}
Our gradient phase-out \newline mechanism demonstrates quantitative improvements over the baseline in the aggregated metrics for 22 states ($12.8\%$ MSE reduction and $15.4\%$ MAE improvement). These improvements are particularly noteworthy given the challenging nature of COVID-19 prediction, where traditional models often struggle with temporal distribution shifts and feature reliability degradation.

The state-specific analysis reveals interesting patterns in performance gains. Arizona and New York show stronger improvements ($16.3\%$ and $17.4\%$ MSE reduction respectively) compared to Massachusetts ($5.7\%$ MAE reduction). However, this comparison should be interpreted carefully, as the data following large data gaps in Arizona and New York were excluded from the in-sample evaluation. The stronger improvements in these states likely reflect the mechanism's effectiveness within the available continuous data segments, rather than its ability to handle missing information directly.

The gradient distribution analysis confirms that our mechanism successfully suppresses testing-related feature gradients as $\alpha$ increases beyond 1, which validates the theoretical design of selective feature suppression. This behavior aligns with the real-world timeline of COVID-19 testing policy changes and demonstrates the model's ability to adapt to evolving data reliability patterns. We note that the actually improvement is likely better than the calculated MSE and MAE reduction, because the possible underestimate of the ground truth data (recovered true cases). 

\subsection{Limitations and Challenges}
Despite the promising results, several limitations deserve careful discussion. We attribute the challenges to two main categories:

\textbf{Data Quality Issues:} The biggest limitation is the wastewater RNA data itself. The concentration of wastewater RNA is subject to many factors and can have sudden changes without any particular reason. We attempted to use weather data to address these sudden changes, but there could be other unaddressed issues. Additionally, approximately $60\%$ of the 13,000 data points contain at least one missing feature, significantly complicating pattern learning. More critically, all states except Massachusetts have some missing wastewater viral RNA data during the period from late 2020 to late 2021, which is the time period when the testing data and recovered confirmed case count was the most reliable. Although we suppressed the dependence of true case count on the wastewater viral RNA in early training data, this data quality problem still inevitably impact the neural network training, as the neural network will treat missing data as a data point with zero viral RNA concentration, which is supposed to be corresponding to a day with zero case load. This prevents the model from fully learning the gradient decay mechanism during this crucial transition period.

\textbf{Model and Feature Engineering Limitations:} Current feature transformations may not capture the underlying data distribution optimally. For example, the log-transform for RNA concentration assumes a log-normal distribution that may not hold across all temporal periods. The neural network architecture, while suitable for demonstrating the gradient phase-out concept, may not be optimal for capturing complex temporal dependencies in epidemiological data. More sophisticated architectures such as temporal convolutional networks or transformer-based models might yield better absolute performance.

\subsection{Discussion and Future Directions}
The gradient phase-out mechanism represents a principled approach to handling temporal feature reliability in predictive modeling. The less a variable is trustworthy in the dataset, the more we should penalize its gradient. By explicitly suppressing the gradient of unreliable variables, this framework offers valuable insights for robust prediction in dynamic real-world environments with varying data quality. The demonstrated improvements across multiple states and the validated gradient suppression behavior provide strong evidence for the approach's effectiveness and theoretical soundness.

Our work contributes to the growing body of research on robust machine learning by improving interpretability under temporal changes in feature reliability. Unlike traditional domain adaptation methods that assume static domains, our approach explicitly models the temporal evolution of feature reliability over time. This perspective is particularly relevant for real-world applications where data collection practices, sensor reliability, or reporting standards change over time. By applying the gradient regularization mechanism to COVID-19 prediction, where feature reliability underwent dramatic changes due to policy shifts and technological adoption, our method demonstrates broad applicability across domains such as sensor networks with known maintenance periods, financial markets during policy changes, or climate modeling with instrument transitions. 

The wastewater RNA data is challenging to process due to many reasons such as unexplained fluctuations and changing measurement methods over time. One potential improvement is to normalize the wastewater SARS-COV-2 viral RNA concentration by that of Pepper mild mottle virus (PMMoV) RNA, because it is consistently detected in untreated wastewater globally, has a high concentration, is highly stable in wastewater, and independent of any human infection \cite{Jafferali2021Benchmarking}. If PMMoV RNA data at wastewater treatment plants becomes available, we will renormalize our wastewater RNA data using the PMMoV benchmark. We expect this to significantly improve the quality of our predictions.

\section*{Acknowledgments}
We would like to thank Biobot Analytics for providing SARS-COV-2 wastewater viral RNA data from 2021-2023, which plays a crucial role in this project. We also thank Kyra Shi and Melita Madhurza for participating in the data processing in the early phase of this project.

\bibliographystyle{siamplain}
\pagebreak
\bibliography{reference}

\end{document}